\documentclass{aa}
\usepackage{graphicx}
\usepackage{txfonts}
\def\p{\partial}

\def\nab{\mbox{\boldmath $\nabla$}}
\def\rb{\bar{\rho}}
\def\kb{\bar{\kappa}}
\def\tb{\bar{T}}
\def\sb{\bar{S}}

\newcommand{\BB}{{\bf B}}

\begin{document}

\title{Magnetic confinement of the solar tachocline}

\author{A. S. Brun\inst{1, 2} \and  J.-P. Zahn\inst{2}}

\offprints{A. S. Brun}

\institute{DSM/DAPNIA/Service d'Astrophysique, CEA Saclay, F-91191 Gif-sur-Yvette, France; AIM, UMR 7158, CEA - CNRS - Universit\'e Paris 7\\
\email{sacha.brun@cea.fr} \and
LUTH, UMR 8102, Observatoire de Paris, F-92195 Meudon, France\\
\email{jean-paul.zahn@obspm.fr} }

\date{Received 25 July 2006; Accepted 26 March 2006}

\abstract{We study the physics of the solar tachocline (i.e. the thin transition layer between differential rotation in the convection zone and quasi uniform rotation in the radiative interior), and related MHD instabilities. }
{We have performed 3-D MHD simulations of the solar radiative interior to check whether a fossil magnetic field is able to prevent the spread of the tachocline.}
{Starting with a purely poloidal magnetic field and a latitudinal shear meant to be imposed by the convection zone
at the top of the radiation zone, we have investigated the interactions between
magnetic fields, rotation and shear, using the spectral code ASH on massive parallel supercomputers.}
{In all cases we have explored, the fossil field diffuses outward and ends up connecting with the convection zone,
whose differential rotation is then imprinted at  latitudes above $\approx 40^\circ$ throughout the radiative interior, according to Ferraro's law of isorotation. Rotation remains uniform in the lower latitude region which is contained within closed field lines. We find that the meridional flow cannot stop the inward progression of the differential
rotation. Further, we observe the development of non-axisymmetric magnetohydrodynamic 
instabilities,  first due to the initial poloidal configuration of the fossil field, and later to the toroidal field produced by shearing the poloidal field through the differential rotation. We do not find dynamo action as such in the radiative interior, since the mean poloidal field is not regenerated. But the instability persists during the whole evolution, while slowly decaying with the mean poloidal field; it is probably sustained by small departures from isorotation.}
{According to our numerical simulations, a fossil magnetic field cannot prevent the radiative spread of the tachocline, and thus it is unable to enforce uniform rotation in the radiation zone. Neither can  the observed thinness of that layer be invoked as a proof for such an internal fossil magnetic field.}

%
\maketitle

\section{Introduction}
 
One of the great achievements of helioseismology was to map the internal rotation of the Sun
(Brown et al. 1989). It revealed that the convection zone rotates differentially, with the angular velocity varying little in depth, and that the radiative interior is in quasi-uniform rotation. The transition between these two regimes occurs in a thin layer, the tachocline, which seems to straddle the boundary between the convection zone and the radiation zone. It is still a matter of debate whether the whole tachocline is turbulent, due to thermal convection, or only part of it. 

Here we shall assume that at least some differential rotation is applied on the top of the radiation 
zone proper, excluding the subadiabatic overshoot region. Then, in the absence of other physical processes, this differential rotation should spread into the radiation zone, as was shown by Spiegel \& Zahn (1992). 
That spread is due to the thermal diffusion of the temperature fluctuation which is generated in latitude, along isobars, by the differential rotation. These are tightly linked by the baroclinic relation;
therefore, as the temperature fluctuation expands downwards, so does the differential rotation.
The process is not just a diffusion in this stably stratified medium, since it also involves meridional currents: 
the temporal variation of all fluctuations  is proportional to the square of the Laplacian, and some refer to it as a hyper-diffusion.
Viscosity too contributes to the spread, but only slightly. The whole process is very slow, but in the present Sun the tachocline 
should reach down to 0.3 in radius, in the absence of any other physical process.

Since this is not observed (Charbonneau et al. 1998; Corbard et al. 1999), one has to invoke a mechanism that tends to 
suppress the differential rotation 
in latitude, which is the cause of the spread. In Spiegel \& Zahn's model, this is achieved by a turbulent 
anisotropic viscosity, with much stronger transport in the horizontal than in the vertical direction, which halts the spread of the tachocline. Established originally in the thin layer limit, this property was confirmed by 2-dimensional 
simulations performed by Elliott (1997). The fact that such turbulence acts to reduce the angular velocity gradient has been 
questioned by Gough \& McIntyre (1998), who cite geophysical examples where horizontal turbulence drives the system away from 
uniform rotation. On the other hand, in Couette-Taylor experiments turbulence reduces the angular velocity gradient, when angular 
momentum increases outward and the flow is linearly stable; in other words it acts to suppress the shear which is responsible for 
the instability (Richard \& Zahn 1999). 
This trend appears also in numerical 3-dimensional simulations of the turbulent overshoot region which have been carried out by Miesch (2003).

Another way to oppose the spread of the tachocline is by magnetic stresses. The first to address the problem were R\"udiger \& Kitchatinov (1997). 
In their model, a steady magnetic field confined in the radiation zone opposes the viscous spread of the differential rotation applied 
on the top; the tachocline reduces then to a very thin Ekman-Hartmann layer.
MacGregor \& Charbonneau (1999) confirmed that result, but they also considered another case, where the poloidal field was allowed to 
thread into the differentially rotating convection zone. A rather weak field then suffices to imprint the differential rotation throughout 
the radiation zone, thus enforcing Ferraro's law 
${\bf B_p}\cdot\nab\Omega=0$, 
where ${\bf B_p}$ is the mean poloidal magnetic field and $\Omega$ the angular velocity  (Ferraro 1937).
These models, however, were rather crude, since they did not allow for 
the Ohmic diffusion of the poloidal field; moreover, they ignored thermal diffusion and the associated meridional circulation.

The model outlined by Gough \& McIntyre (1998)  includes most  physical processes that are likely to operate in the upper radiation zone: 
the tachocline circulation is driven by thermal diffusion, and its spread is halted by the fossil field that resides beneath. The field 
does not penetrate into the tachocline, except perhaps in a narrow region of up-welling flow, and therefore it does not transmit the 
differential rotation of the convection zone to the deep interior. Actual calculations of this model have been carried out by 
Garaud (2000, 2002), directly seeking the stationary solution sketched by Gough \& McIntyre, although in her case the tachocline 
circulation is driven through Ekman pumping; some of her solutions show a tendency for magnetic confinement, but a substantial 
differential rotation persists throughout the radiation zone, particularly at high latitude.

The models discussed above favor a slow version of the tachocline with a typical ventilation time of the order of  $10^{6}$ yr.  Gilman and collaborators (Gilman \& Fox 1997; Dikpati, Cally \& Gilman 2004 and references therein) developed a series of models that showed that the tachocline could become unstable through
magnetic instability of toroidal structures embedded in it, resulting
in a fast latitudinal angular momentum transport that suppresses the shear and limits its inward 
diffusion. Forgacs-D\'ajka and Petrovay (2002) considered the effect of an oscillating dynamo field of 22 yr period and found that it too
could suppress the spread of the solar tachocline. Both models favor a rather fast tachocline with ventilation times of the order of a rotation period or the duration of the solar cycle (Gilman 2000).

In the present paper we do not deal with such short timescales, and we examine a model of the solar radiative zone and tachocline similar to that of Gough \& McIntyre. But we 
consider the time-dependent problem, starting from various initial conditions, for two reasons: i) both the tachocline spread and the diffusion of the magnetic field proceed very slowly, and thus it is not clear 
that a stationary solution can be reached by the age of the Sun; ii) such a stationary solution may not be unique, but may actually depend 
on the initial conditions. A similar approach has been taken by Sule et al. (2005); we differ from their work in that we let the 
poloidal field diffuse, and in our case the tachocline circulation is driven mainly by thermal diffusion, as in the Sun.

Moreover, we use for our simulations the 3-dimensional code ASH (see further down) and resolve the Alfv\'en crossing time; this enables us to describe non-axisymmetric 
MHD instabilities which could lead to drastic reconfigurations of the magnetic field, as was recently demonstrated numerically 
by Braithwaite and Spruit (2004; see also Braithwaite \& Nordlund 2005), following the pioneering work of Tayler (Tayler 1973; Pitts \& Tayler 1985). 

\section{The model}
We make use of the well-tested hydrodynamic ASH code (Anelastic Spherical
Harmonic; see Clune et al. 1999; Miesch et al. 2000; Brun \& Toomre
2002) which was originally designed to model the solar convection zone. It has since been extended to include the magnetic induction equation and the feedback of
the field on the flow via Lorentz forces and Ohmic heating (see Brun, Miesch \& Toomre 2004 for
more details).  Here we apply it to examine the nonlinear interaction between a shearing flow, i.e. the
differential rotation imposed by the convection zone on the top of the radiative interior,
and a fossil magnetic field. Our goal is to assess whether such a field can hinder the spread
of the tachocline deep into the radiation zone and thus keep it thin, as was revealed by 
helioseismic inversions (Charbonneau et al. 1998; Corbard et al. 1999). 

Our numerical model is a simplified portrayal of the solar radiation
zone: solar values are taken for the heat flux, rotation rate, mass
and radius, and a perfect gas is assumed.  The computational domain
extends from $r_{\rm bot}=0.35 \, R_\odot$ to $r_{\rm top} = R = 0.72 \, R_\odot$ ($R_\odot$ is the solar radius); we thus focus on the bulk of the stably stratified zone, excluding  the nuclear central region, and we ignore its possible back reaction on the convective envelope.
The vertical extent of the computational domain is therefore $D=2.5\times 10^{10}$ cm and 
the background density varies across the shell by about a factor of 17.  

\subsection{Governing equations}

The ASH code solves the full set of 3--D MHD anelastic equations of motion (Glatzmaier 1984) in a rotating spherical shell 
on massively-parallel computer architectures (Brun, Miesch \& Toomre 2004). 
Depending on the sign of the initial mean entropy gradient considered in the model, either negative or positive, convection or radiation zones can be simulated.
The anelastic approximation is used to retain the important effects of density stratification without 
having to track the sound waves (i.e. $\p \rho/\p t=0$). 
The MHD anelastic equations are fully nonlinear in velocity and magnetic field variables, but under 
the anelastic approximation the thermodynamic variables are linearized with respect to a spherically
symmetric and evolving mean state, with density $\rb(r,t)$, pressure
$\bar{P}(r,t)$, temperature $\tb(r,t)$ and specific entropy $\sb(r,t)$.  Fluctuations
about this mean state are denoted by $\rho$, $P$, $T$, and $S$.  The
resulting equations are:
\begin{eqnarray}
\nab\cdot(\rb {\bf v}) &=& 0, \\
\nab\cdot {\bf B} &=& 0, \\
\rb \left(\frac{\p {\bf v}}{\p t} + ({\bf v}\cdot\nab){\bf v}+2{\bf \Omega_0}\times{\bf v}\right) 
& = & -\nab P + \rho {\bf g} + \frac{1}{4\pi} (\nab\times{\bf B})\times{\bf B} \nonumber \\
&-& \nab\cdot\mbox{\boldmath $\cal D$}-[\nab\bar{P}-\rb{\bf g}], \\
\rb \tb \frac{\p S}{\p t}+\rb \tb{\bf v}\cdot\nab (\sb+S)&=&\nab\cdot[ \rb c_p (\kb \nab \tb
+  \kappa \nab T)]\nonumber \\
&+&+2\rb\nu\left[e_{ij}e_{ij}-1/3(\nab\cdot{\bf v})^2\right] \nonumber \\
 &+& \frac{4\pi\eta}{c^2}{\bf j}^2 + \rb \bar{\epsilon},\\
\frac{\p {\bf B}}{\p t}&=&\nab\times({\bf v}\!\times \!{\bf B})-\nab\times(\eta\nab\!\times\!{\bf B}),
\end{eqnarray}
where ${\bf v}=(v_r,v_{\theta},v_{\phi})$ is the local velocity in spherical coordinates in 
the frame rotating at constant angular velocity ${\bf \Omega_0}$, ${\bf g}$ is the 
gravitational acceleration, ${\bf B}=(B_r,B_{\theta},B_{\phi})$ the magnetic field, 
${\bf j}=c/4\pi\, (\nab\times{\bf B})$ the current density, and
$c_p$ the specific heat at constant pressure.
$\mbox{\boldmath $\cal D$}$ is the viscous stress tensor, involving the components
\begin{eqnarray}
{\cal D}_{ij}=-2\rb\nu[e_{ij}-1/3(\nab\cdot{\bf v})\delta_{ij}],
\end{eqnarray}
where $e_{ij}$ is the strain rate tensor. 

The radiative diffusivity $\kb$ applied to the mean temperature gradient $\nab \tb$ takes the solar value, hence the radiation flux is that of the Sun to keep the model consistent. But the diffusion coefficient $\kappa$ operating on the temperature fluctuations will be enhanced in our simulations, as explained below, and so also will the viscosity $\nu$ and the Ohmic diffusivity $\eta$.  
A volume heating term $\rb \bar\epsilon$ is also included in these equations for completeness (see Brun, Browning \& Toomre 2005). 

Finally we use the linearized equation of state, assuming the ideal gas law:
\begin{equation}
\frac{\rho}{\rb}=\frac{P}{\bar{P}}-\frac{T}{\tb}=\frac{P}{\gamma\bar{P}}-\frac{S}{c_p},
\end{equation}
where $\gamma$ is the adiabatic exponent.  We neglect the composition gradient due to element settling. The reference or mean state, which is
indicated by overbars, is derived from a one-dimensional solar
structure model (Brun et al.\ 2002); it is frequently updated with
the spherically-symmetric components of the thermodynamic fluctuations
as the simulation proceeds.  The model begins in hydrostatic balance so the
bracketed term on the right-hand-side of equation (3) initially
vanishes. Departures from this hydrostatic balance are found to be small.

\subsection{Integration domain, boundary conditions}

The system of equations to be solved with ASH requires 12 boundary conditions.  At 
the top  of the numerical domain we impose a differential rotation meant to be enforced by the turbulent convection zone. 
Following helioseismic inversions (Thompson et al. 2003), we apply 
the following law:
\begin{equation}
\Omega(r_{\rm top}, \theta)= A + B \cos ^2 \theta + C \cos ^4 \theta .
\label{diffrot}
\end{equation}
The uniform rotation which has the same specific angular momentum (Gilman et al. 1989) is
\begin{equation}
\Omega_0 = A + {1 \over 5} B + {3 \over 35} C ;
\label{uniformrot}
\end{equation}
we shall apply this uniform rotation initially to the whole radiation zone, and therefore
there will be no net viscous flux across the boundary.

For the remaining top and bottom boundary conditions, we impose:
\begin{enumerate}
\item Impenetrable walls at the top and bottom: 
\[
v_r=0|_{r=r_{\rm bot},r_{\rm top}},
\]
\item Stress free at the bottom: 
\[
\frac{\p}{\p r}\left(\frac{v_{\theta}}{r}\right)=\frac{\p}{\p r}\left(\frac{v_{\phi}}{r}\right)=0|_{r=r_{\rm bot}},
\]
\item Constant entropy gradient at the top and bottom: 
\[
\frac{\p \sb}{\p r}=cst|_{r=r_{\rm bot},r_{\rm top}} ,
\]
\item We match the magnetic field to an external potential field at the top and bottom: 
\[
\BB=\nab \Phi \rightarrow \Delta \Phi =0|_{r=r_{\rm bot},r_{\rm top}}.
\]
\end{enumerate}
Our choice of magnetic boundary conditions ensures that no magnetic torques are applied to the numerical domain.

\subsection{Choosing the physical parameters}

In Gough \& McIntyre's model, a magnetic boundary layer, or magnetopause, separates the magnetic interior from the tachocline 
void of magnetic field; its thickness $\delta$ is determined by the relative strength of the two diffusion processes and that of the magnetic field:
\begin{equation}
\left({\delta \over R}\right)^6 = 
 {2 \over L^4} \, {(t_{\rm A})^2 \over t_{\rm ES} \, t_{\rm B}} ,
 \label{smalldelta}
\end{equation}
 where $R=r_{\rm top}$, and $L \approx 4.5$ is related to the latitudinal structure of the tachocline. Here we have introduced
 the Alfv\'en crossing time $t_{\rm A}= R / V_{\rm A} = R \sqrt{4\pi \rb} / B $; $ t_{\rm B} = R^2 / \eta$ is the magnetic diffusion time, 
and $t_{\rm ES} = (N / \Omega_0)^2 (R^2 / \kappa)$ the local Eddington-Sweet time.
$B$ is the magnetic field strength and $N$ the buoyancy frequency.

 Expressing that the diffusion of the magnetic field is halted by the downflow of the tachocline circulation, 
 whose ventilation time is given by
 \begin{equation}
 t_{\rm vent} = L \left({\Delta \over R}\right)^4 \left({\widehat \Omega \over \Omega_0}\right)  t_{\rm ES} \label{vent} ,
\end{equation}
one obtains the following relation 
between the thickness $\delta$ of the magnetopause and  $\Delta$, that of the tachocline :
\begin{equation}
\left({\Delta \over R}\right)^3 = {1 \over L} \left({\delta \over R}\right) {t_{\rm B} \over t_{\rm ES}} \left({\widehat \Omega \over \Omega_0}\right),
\label{largedelta}
\end{equation}
where  $\widehat \Omega / \Omega_0 \approx 1/5$  is a measure of the differential rotation.
 
\begin{figure}[ht]
\centering
\includegraphics[width=1.0\linewidth]{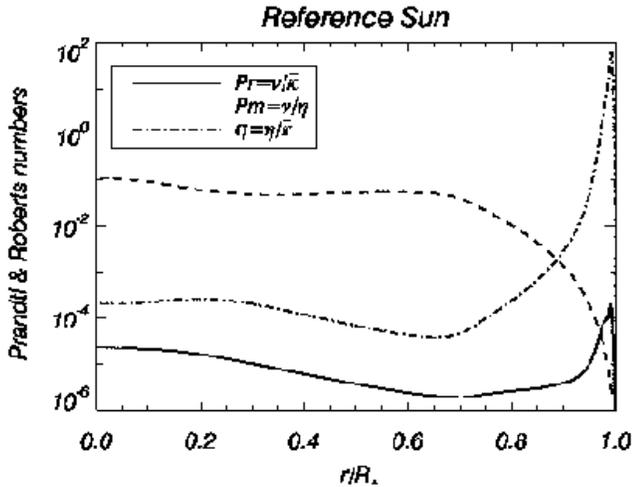}
\caption{\label{diff-coeff} Radial dependence of the Prandtl $P_r$, magnetic Prandtl $P_m$ and Roberts $Q$ numbers in the Sun based
on the microscopic values of the diffusivities for a ionized hydrogen gas (Zeldovich et al. 1983).}
\end{figure}
 
We note that the size of these boundary layers does not depend too sensitively on the value of the diffusivities and on the strength of the magnetic field, which means that we will not alter too much the problem when we increase these values to meet the numerical requirements.  Figure \ref{diff-coeff} shows the variation with depth of the non-dimensional numbers that characterize the microscopic diffusivities in the Sun, and typical values for the relevant parameters in the upper radiation zone are given in Table~\ref{tble}. Due to limitations in computing resources, no simulation achievable
now or in the near future can hope to directly use microscopic values
for the diffusivities. In our model we had to increase all
 diffusivities to be able to resolve the smallest scales (i.e. the thickness of the Ekman layer, and the size of the turbulent eddies induced by the instabilities described later in \S \ref{instab}), but we took care to respect their hierarchy. We chose the viscosity $\nu$ as small as allowed by our numerical resolution, in order 
to ensure that the tachocline be driven by thermal diffusion, as in the Sun, rather than through Ekman pumping. Another way to look at the problem is 
to make sure that the spread due to thermal hyper-diffusion, which evolves as $\Delta/R \approx (t/t_{\rm ES})^{1/4}$, still exceeds the viscous spread 
$\Delta/R \approx (t/t_\nu)^{1/2}$ at the age of the Sun (with $t_{\nu}=R^2/\nu$), which measured in Eddington-Sweet time is $t_\odot/t_{\rm ES}= 1.2 \, 10^{-3}$; this is guaranteed 
if the Prandtl number is less than $\nu/\kappa = 3 \, 10^{-3}$, hence our choice of $\nu/\kappa = 10^{-3}$.

 The Froude number $\Omega_0 /N$ was also increased to meet the numerical requirements; since this enhances the ventilation speed, it may also account for the fact that the Sun was rotating more rapidly
in the past. With these parameters, taking a field of 1kG yields a tractable thickness for the magnetic boundary layer: $\delta /R = 2.2 \, 10^{-2}$ according 
to (\ref{smalldelta}). From (\ref{largedelta}) the corresponding tachocline thickness is then $\Delta /R = 2.1 \, 10^{-2}$, and the magnetic Reynolds number based on the      Alfv\'en speed
$R_m = t_B/t_A$ will reach a few $10^3$ in the computational domain.
\begin{table}
\begin{center}
\begin{tabular}{|*{4}{c|}} 
   \hline 
   Parameter & Symbol & Sun & Simulation  \\ 
   \hline
  mean thermal diffusivity & $\bar{\kappa}$ & $10^7$ & $10^{7}$ \\      
  thermal diffusivity & $\kappa$ & $10^7$ & $8 \, 10^{12}$ \\
    magnetic diffusivity & $\eta$ & $10^3$ & $8 \, 10^{10}$ \\
    viscosity & $\nu$ & $30$ & $8 \, 10^{9}$ \\
       \hline  
        Prandtl number & $\nu/\kappa$ & $3\, 10^{-6}$ & $ 10^{-3}$ \\
       Roberts number & $\eta/\kappa$ & $10^{-4}$ & $ 10^{-2}$ \\
        magnetic Prandtl nb & $\nu/\eta$ & $3\, 10^{-2}$ & $ 10^{-1}$ \\
       Froude number & $(\Omega_0 / N)^2$ & $2 \, 10^{-6}$ & $ 10^{-4}$ \\
      Ekman number & $\!\nu /2\Omega_0 R^2\!$ & $2 \, 10^{-15}$ & $ 6 \, 10^{-7}$ \\
         \hline 
       rotation frequency & $\Omega_0$ & $ 3 \, 10^{-6}$ & $ 3 \, 10^{-6}$ \\ 
       Alfv\'en time (for 1 G) & $t_{A}$ & $5 \, 10^{10}$ & $5 \, 10^{10}$ \\
    Ohmic diffusion time & $t_{B}$ & $2.5 \, 10^{18}$ & $ 3.125 \, 10^{10}\!$ \\
       viscous diffusion time & $t_{\nu}$ & $\!8.333 \, 10^{19}\!$ & $ 3.125 \, 10^{11}\!$ \\
    Eddington-Sweet time & $t_{ES}$ & $1.25 \, 10^{20}$ & $ 3.125 \, 10^{12}\!$ \\
   \hline 
\end{tabular}
\bigskip 
  \caption{Typical values of the relevant parameters in the upper radiation zone of the Sun, and values adopted for the model  at 
  $r_{\rm top} = 0.72 R_\odot $ (in cgs units, when applies). 
In the simulation, the thermal diffusivity $\bar\kappa$ operating on the mean temperature gradient takes its solar value; only the diffusivity $\kappa$ applied on the temperature fluctuations is increased.}
 \label{tble}
   \end{center} 
\end{table}

\subsection{Numerical resolution}

The global code ASH solves in a 3-D rotating spherical shell the set of equations (1-5) 
in the anelastic approximation. This has the great advantage that the Courant stability criterion
is based on the slow motions present in the radiative zone rather than on the sound speed, 
thus allowing for much larger time steps. In the MHD context, the anelastic approximation filters 
out fast magneto-acoustic waves but retains the Alfv\'en and slow magneto-acoustic modes.

The ASH code is based on the so-called pseudo-spectral numerical method to solve the
system of partial differential equations introduced above (Glatzmaier 1984; Canuto et al. 1995; Boyd 1989; Clune et al. 1999).
In this numerical method, the main variables are projected onto orthogonal functions/polynomials rather than discretized
on a mesh grid as with finite differences. The advantage of this method is that in spectral space, derivatives are
straightforward multiplications,  rather than variations as in a finite difference scheme, which are numerically less precise. 
Since we want to solve the MHD equations in spherical geometry, it is natural to
use the spherical system of coordinates ($r, \theta, \phi$) and to project the main variables on spherical 
harmonics $Y_{\ell m}(\theta,\phi)$. This approach has the advantage that the
spatial resolution is uniform everywhere on a sphere when a complete
set of spherical harmonics is used up to some maximum in degree $\ell$
(retaining all azimuthal orders $m \leq \ell$ in what is known as triangular truncation). Gaussian quadrature
imposes that we use the relation $\ell_{max}=(2 N_{\theta} - 1) / 3$, and $N_{\phi} = 2 N_{\theta}$ for the horizontal resolution (Boyd 1989). 
For the radial direction we choose to project our functions onto Chebyshev polynomials $T_n(r)$, using $N_r$
Gauss-Lobatto colocation points (Canuto et al. 1995). Typical grid resolution for our models is $N_r=193$, $N_{\theta}=128$ and $N_{\phi}=256$; the time-step was chosen such as to resolve the gravity modes. 

In order to ensure that the mass flux and the magnetic field
remain divergence-less to machine precision throughout the simulation,
we use a toroidal--poloidal decomposition:
\begin{eqnarray}
{\rb\bf v}=\nab\times\nab\times (W \hat{\bf e}_r) +  \nab\times (Z \hat{\bf e}_r), \\ 
{\bf B}=\nab\times\nab\times (C \hat{\bf e}_r) +  \nab\times (A \hat{\bf e}_r) ~~~. 
\end{eqnarray}
More details on the numerical method can be found in Clune et al. (1999) and Brun, Miesch \& Toomre (2004).

\subsection{Initial conditions}

At $t=0$  the computational domain is motionless, except for a uniform rotation
$\Omega = \Omega_0$ defined in (\ref{uniformrot}), and for the differential rotation imposed at the top boundary
$r = r_{\rm top}$, where the following values have been chosen for  $A$, $B$ and $C$ in expression (\ref{diffrot}): $A/2 \pi=456$, $B/2 \pi =-42$ and $C/2\pi=-72$ nHz. 

Initially  we apply a purely axisymmetric meridional field $\vec B_p = B_r \vec e_r + B_\theta \vec e_\theta$, with
\begin{equation}
B_r=  {B_0 \over r^2 \sin \theta} {\partial \Psi \over \partial \theta} , 
\qquad
B_\theta = - {B_0 \over r \sin \theta} {\partial \Psi \over \partial r}  ;
\end{equation}
$\Psi (r, \theta)$  is constant on field lines, and  is related to the azimuthal component of the vector potential: $\Psi = r \sin \theta A_\phi$. 
For a dipolar field $\Psi \propto \sin ^2 \theta \, R^3 / r$; here we take instead
\begin{eqnarray}
&\Psi =  \left({r / R}\right)^2 (R_B - r)^2 \sin^2 \theta  \; &\quad {\rm for} \; r \leq R_B  \nonumber \\
&\Psi = 0  \; &\quad {\rm for}\; r \geq R_B ,
\end{eqnarray}
which confines the field in the sphere of radius $R_B$.\footnote{Charbonneau and MacGregor (1993) took a similar form for $\Psi$; we differ from them 
by enforcing the continuity of $B_\theta$ at $R_B$, thus suppressing there the current sheet.} 

This parameter $R_B$ thus determines  the degree of confinement of the initial field; we shall consider the 3 cases:
\begin{itemize}
\item
I. $R_B \gg r_{\rm top}$ - the field threads into the convection zone,
\item
II. $R_B = r_{\rm top}=0.72 R_\odot$ - the field is just barely confined in the radiation zone,
\item
III. $R_B = 0.64 R_\odot$ - the initial field is buried deep in the radiation zone.
\end{itemize}
   
\section{Results}

The interaction of rotation and magnetic field in
the solar radiative zone is found to be very sensitive to the initial 
magnetic field configuration, especially in the early phases of the temporal evolution; 
we shall examine in turn the cases I, II and III defined above.
To ease the comparison with the real Sun, we have rescaled the time by the ratio $t_{\rm ES} {\rm(Sun)}/t_{\rm ES} {\rm(simulation)} = 4 \, 10^7$; thus from here on we quote the time in that `solar equivalent unit'.

\subsection{Case I: Initial magnetic field threading into the convection zone}

\begin{figure*}[ht]
\centering
\includegraphics[width=0.75\linewidth]{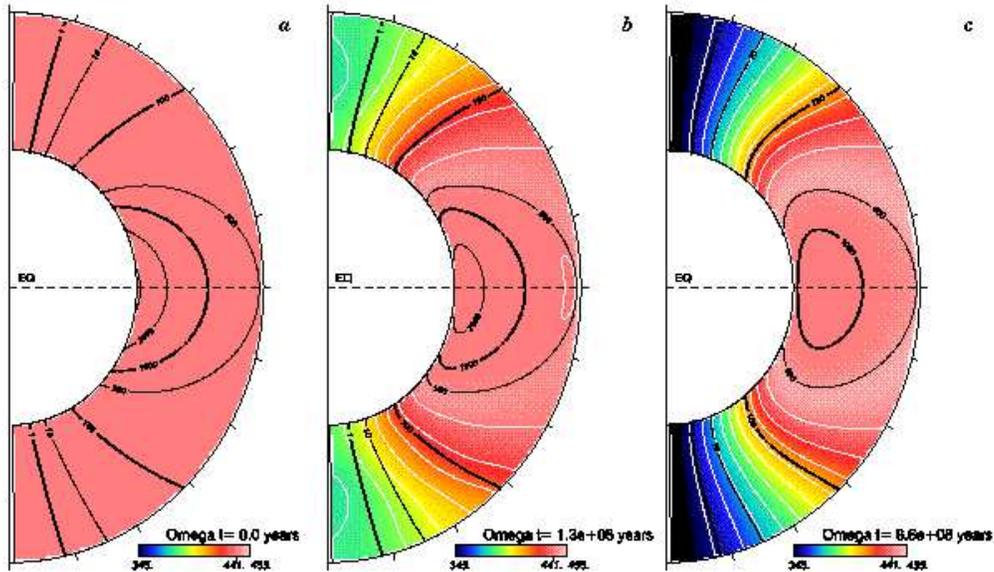}
\caption{\label{case1} Case I. Temporal evolution, over 860 Myr in equivalent solar units, of the angular velocity $\Omega$ (color contours and white lines) and the mean axisymmetric 
poloidal field (superimposed black lines, whose labels refer to the field strength in the equatorial plane). The time is measured in solar equivalent units, i.e. rescaled by the ratio $t_{\rm ES} {\rm(Sun)}/t_{\rm ES} {\rm(simulation)}
= 4 \, 10^7$. Initially, the field already penetrates into the convection zone; it quickly imprints the differential rotation imposed at the top on the whole radiation zone.}
\end{figure*}

In this first case, at $t=0$ the magnetic field lines are already threading into the
differentially rotating convection zone, as shown
in Fig.~\ref{case1}a. The torque applied by the shear at
the top of the domain is quickly transmitted throughout the radiation zone, first at high latitudes 
where the direction of the field lines is almost radial (Fig.~\ref{case1}b) 
and then more slowly at lower latitudes (Fig. ~\ref{case1}c).
Such a strong differential rotation in latitude in the solar radiative interior 
is not observed in helioseismic inversions (Thompson et al. 2003; Couvidat et al. 2003).
Clearly, in this case the magnetic stresses do not oppose the 
inward propagation of the shear but they favor it. (Similar results have been
found in 2-D by MacGregor \& Charbonneau 1999.)
The transport of angular momentum by Maxwell stresses is very fast, and  
prevents the magnetopause from forming. 
Meridional flows do exist but they are not found to play a significant role: they are unable to oppose the spread of the latitudinal shear. 
Their amplitude in the simulation is about 1~m/s, which corresponds (after rescaling) to a real solar ventilation speed of 10$^{-7}$ m/s, in excellent agreement with the theoretical prediction (Gough \& McIntyre 1998).

The interaction between the poloidal field and the shear leads
initially to a fast increase of the axisymmetric toroidal field, which then saturates as the
isocontours of $\Omega$ become more and more aligned with the 
axisymmetric poloidal field, tending to Ferraro's law of isorotation. 
Its strength, over a short interval of time, is larger than that of the axisymmetric
poloidal field, before equilibrating at  about the same amplitude. 
These mean toroidal fields are found to become unstable to a $m=1$ instability.  Deeper down and in the early phase of evolution, we observe the fast grow of high-$m$ non-axisymmetric perturbations, which is due to
a MHD instability of the purely poloidal field configuration considered. Both the low and high-$m$ instabilities will be described in more detail in \S \ref{instab}.

\subsection{Case II: Initial field barely confined in the radiation zone}
\begin{figure*}[ht]
\centering
\includegraphics[width=0.75\linewidth]{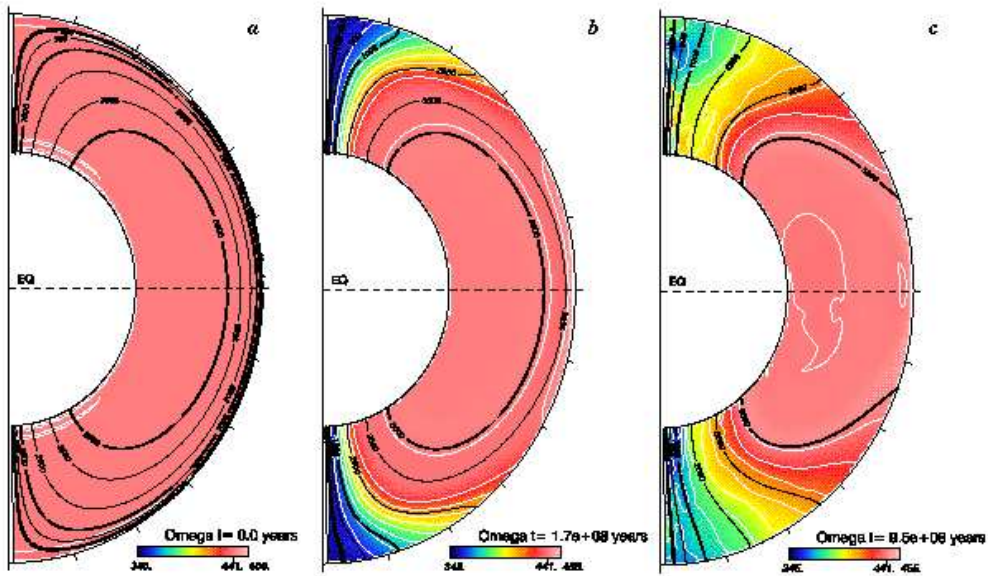}
\caption{\label{case2} Case II. Temporal evolution, over 950 Myr in solar equivalent units, of the angular velocity $\Omega$ (color contours and white lines) 
and the mean axisymmetric poloidal field (black lines). The initial field does not penetrate into the convection zone, but the outcome is the same as in case I.}
\end{figure*}

In the second case, the magnetic field lines do not penetrate into the
convection zone but they come very close to it, as can be seen in Fig. \ref{case2}a. 
At the beginning of the simulation,  the viscous torque associated with the imposed shear at the top of the domain 
does not drag the magnetic field lines, and as a consequence no physical processes are available to quickly transmit 
the latitudinal shear. This particular configuration 
of the magnetic field has been chosen in the 2-D calculations of R\"udiger \& Kitchatinov (1997) and
MacGregor \& Charbonneau (1999) to demonstrate the confinement of the tachocline. 
If the initial poloidal magnetic field is kept fixed
in time, as in their calculations, its peculiar topology is quite efficient at stopping the inward diffusion of the latitudinal shear. 
However,  in our simulations the
magnetic field is allowed to diffuse outward, where it connects with 
the imposed rotation (Fig. \ref{case2}b). As the more intense field lines reach the top
boundary, and slowly drift toward the polar regions while becoming more radial, 
isorotation is established here also, first at
high latitude and then at lower latitude (Fig. \ref{case2}c). It takes about 8 times longer for 
Ferraro's law to be achieved in case II than in case I. However this is still a fast process, 
accomplished at an age of 0.2 Gyr in equivalent solar units.
Here again the internal magnetic field fails to prevent the
inward spread of the shear because the magnetopause does not have the time to form, and 
as a consequence there is no meridional circulation to oppose the spread of the latitudinal shear
by  horizontally bending the magnetic field lines.
Meridional flows do exist but here again they do not play a significant role. 
\begin{figure*}[!ht]
\center
\includegraphics[width=1.0\linewidth]{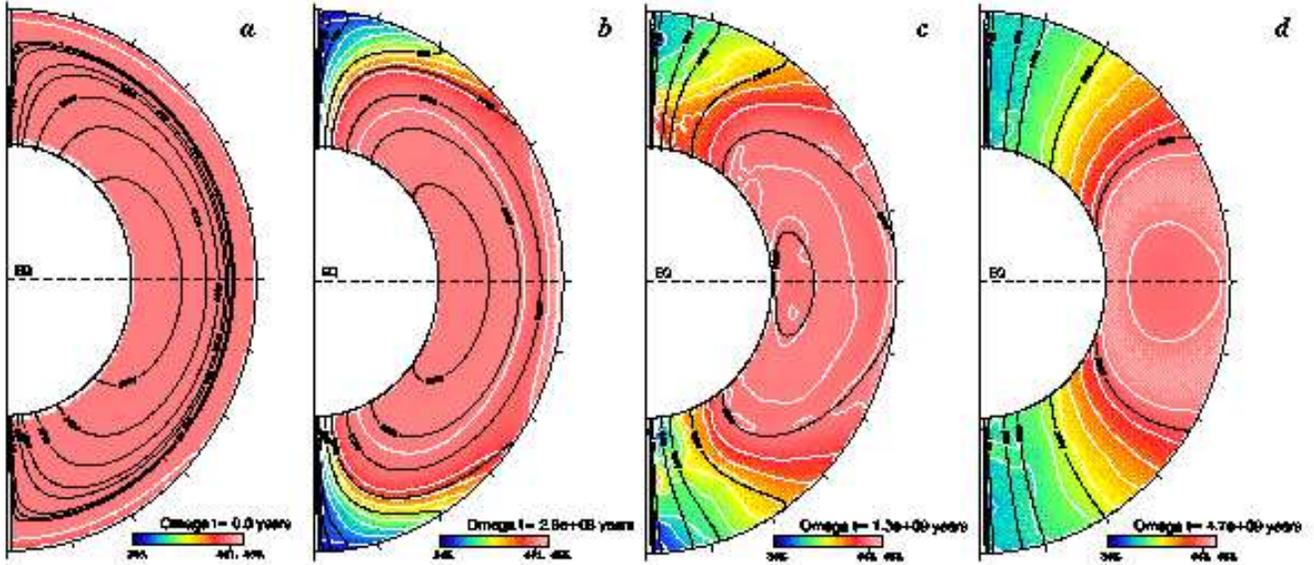}
\caption [] {\label{figOmT} Case III. Temporal evolution of the angular
velocity $\Omega$ (color contours and white lines) and the mean axisymmetric poloidal field (superimposed black lines), which initially is buried below the convection zone. 
{\bf a-d)} sequence spanning 4.7 Gyr (in solar equivalent units) of the dipolar magnetic field in the presence of rotation and shear. 
The field lines gradually connect with the imposed top shear, thereby enforcing differential rotation at all depths above latitudes greater than $\approx 40^\circ$, at the (equivalent) age of the Sun.
Note also in frames {\bf c-d} the distortion  near the poles of the lines of constant $\Omega$, which is associated with the Tayler instability.}                  
\end{figure*}
\begin{figure*}[!ht]
\center
\includegraphics[width=1.0\linewidth]{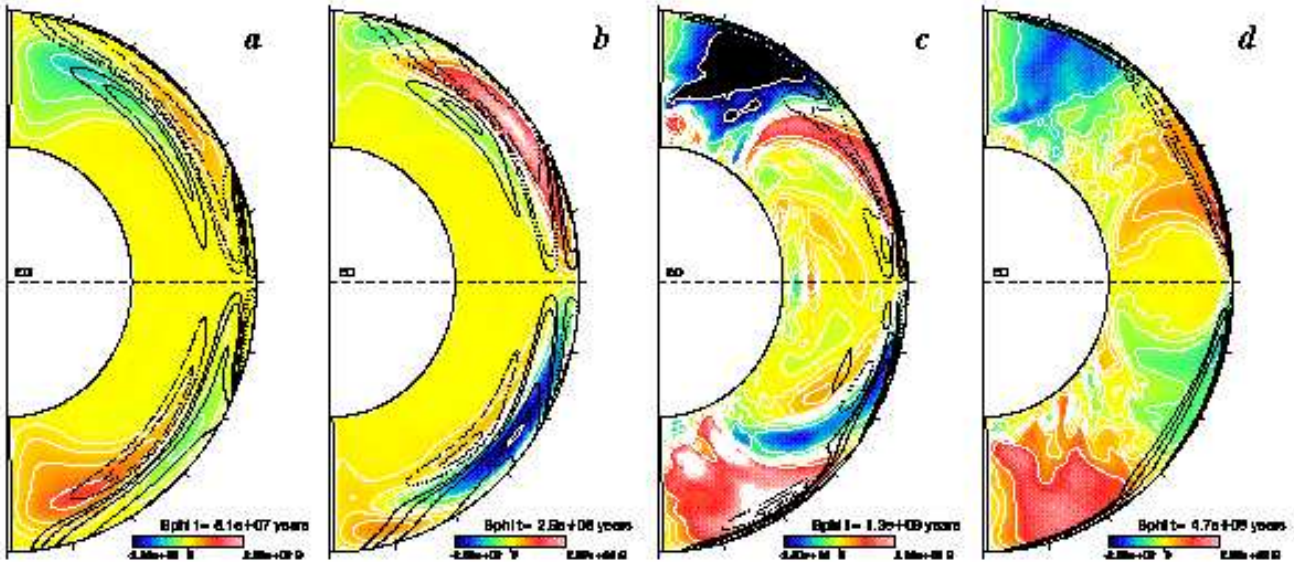}
\caption[]{\label{figMC} {\bf a-d)} Case III. Temporal evolution of the associated toroidal field (color contours) and of the meridional circulation (superimposed black lines) at t= 81, 280 Myr, 1.3 and 4.7 Gyr (solar equivalent units). 
We clearly see the early formation  of a magnetopause with an intense mean (axisymmetric) toroidal field, and then
the development at high latitudes of a toroidal field associated with the progression of the latitudinal shear into the radiation zone. Note the highly distorted contours in {\bf c-d}, a sign that the Tayler instability has reached high amplitude,
and the strongly radially squeezed counter-cells near the surface at low latitudes.
}                  
\end{figure*}

 As soon as the field lines  make contact with the shear, the 
situation we observed in the first case is again realized, but the toroidal field takes about 8 times
longer to reach the same level as in case I. The energy in the mean toroidal field saturates as the
isocontours of $\Omega$ become more and more aligned with the 
axisymmetric poloidal field. As in case I, we find that the mean toroidal field becomes unstable to a $m=1$ 
instability at high latitude, and that a rapid non-axisymmetric MHD instability 
of the mean poloidal field develops in the early phase. The timing of these instabilities varies between the
3 cases, being the strongest for case III (see below \S 4).
We can thus conclude that case II leads to the same state of differentially rotating radiative
interior as case I, except that this occurs later.  Hence we do not confirm the results of 
R\"udiger \& Kitchatinov (1997) and MacGregor \& Charbonneau (1999), that a barely confined field
is able to prevent the spread of the tachocline. The temporal evolution of the mean poloidal
field plays  a crucial role here by allowing the field to connect to the imposed upper shear.


\subsection{Case III. Initial field buried deeply in the radiation zone}
In this last case, the magnetic field lines are initially confined below $r=R_B= 0.64R_\odot$, to let the tachocline 
penetrate somewhat into the interior before encountering the fossil field (Fig. \ref{figOmT}a).
When the field lines make contact with the shear (see Fig. \ref{figOmT}b), we notice a fast increase of the 
mean toroidal energy (TME) in a thin band extending over a large range in latitude, which corresponds to the magnetopause anticipated by Gough \& McIntyre (see Fig. \ref{figMC}a). 
In case III, it takes longer for the TME to reach a level comparable to that of cases I and II, as  
expected since at the beginning the mean poloidal field does not interact with the shear. 

At 280 Myr, the meridional circulation in the tachocline consists of three superposed counter-rotating cells in each hemisphere, of which the largest spans all latitudes,  like a dipole. The pattern resembles that obtained by Elliott (1997) in the non-magnetic case. Here also the circulation is octopolar close  to the top boundary, with a tilted up-welling located around the latitude of 35$^\circ$.  This meridional circulation is unable to prevent the outward diffusion of the magnetic field lines, even in the down-dwelling regions (Figs. \ref{figOmT}b-d); it does not succeed in squeezing the field lines in narrow up-wells so as to imprint below only a small amount of differential rotation. The direct consequence of this inefficiency is that the field lines eventually connect to a large portion of the 
shear imposed at the top of the domain by diffusing through the magnetopause.
This results in an upward shift of the magnetopause, which is the site of a  strong generation of toroidal field (via the $\Omega$-effect) - notice the migration of $B_{\phi}$ between Figs. \ref{figMC}a and b. Then as the field lines
gradually connect to the shear imposed at higher latitudes, becoming more radially oriented, 
they communicate the differential rotation of the convection zone to all depths via Maxwell stresses, as in the two previous cases. At 4.7 Gyr, the meridional circulation has shrunk into two very thin dipolar cells in each hemisphere, one being hardly distinguishable near the equator, and the differential rotation has invaded all latitudes above 40$^\circ$  from top to bottom.

We have run Case III with different values of the confinement depth $R_B$ and of the Prandtl number $\nu/\kappa$, which allows us to conclude that the speed at which the differential rotation penetrates into the radiative interior depends mainly on the time it takes for the poloidal field
lines to connect with the convection zone: the deeper the field is buried initially and the longer it can maintain uniform rotation within its closed field lines, while allowing for a thicker tachocline. Thus Ferraro's law of isorotation, with a radiative interior rotating differentially, is the
most likely outcome of the interaction of the poloidal field with the latitudinal shear. The
initial state of solid rotation only lasts as long as the fields lines have not
diffused outward and connected with the convection zone.

\begin{figure}[!ht]
\center
\includegraphics[width=1.0\linewidth]{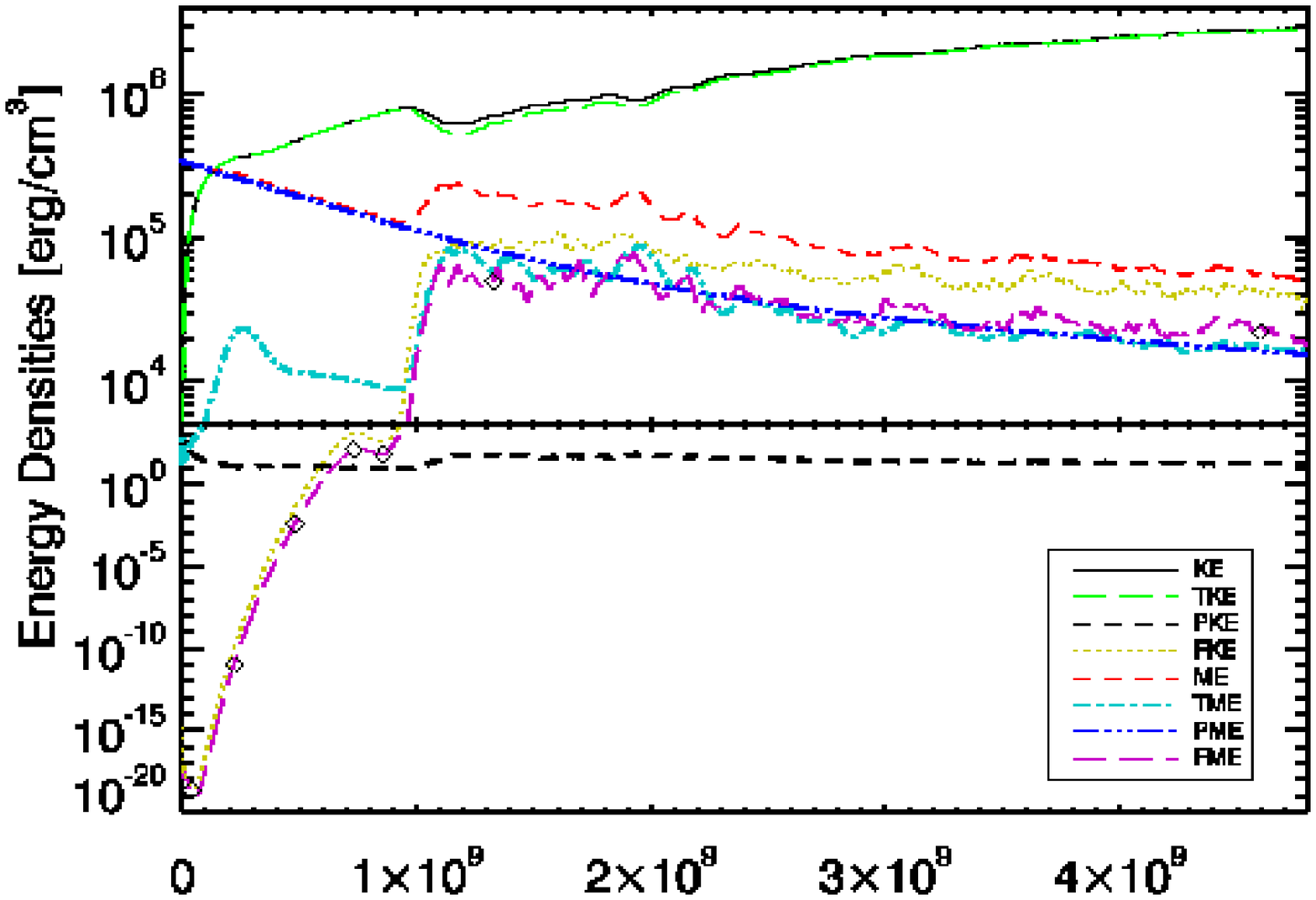}
\caption[]{\label{figKEMETac} Case III: temporal evolution of kinetic and magnetic energies in the
presence of rotation and shear. We clearly see the 2 phases of non-axisymmetric instability (characterized by
the fast increase of FKE and FME): the first, for $t < 1$ Gyr, is associated with the unstable configuration of the purely dipolar 
initial field, and the second, for $t >1$ Gyr, with the toroidal field produced by winding up the poloidal field through the differential rotation. The time is given in solar equivalent units, i.e. rescaled by the ratio $t_{\rm ES} {\rm(Sun)}/t_{\rm ES} {\rm(simulation)}$. Diamonds have been superimposed on the FME curve to indicate the temporal sampling used in Figs. \ref{figShTac} and \ref{spectrum-bottom}.}                  
\end{figure}

\begin{figure*}[!ht]
\center
\includegraphics[width=0.85\linewidth]{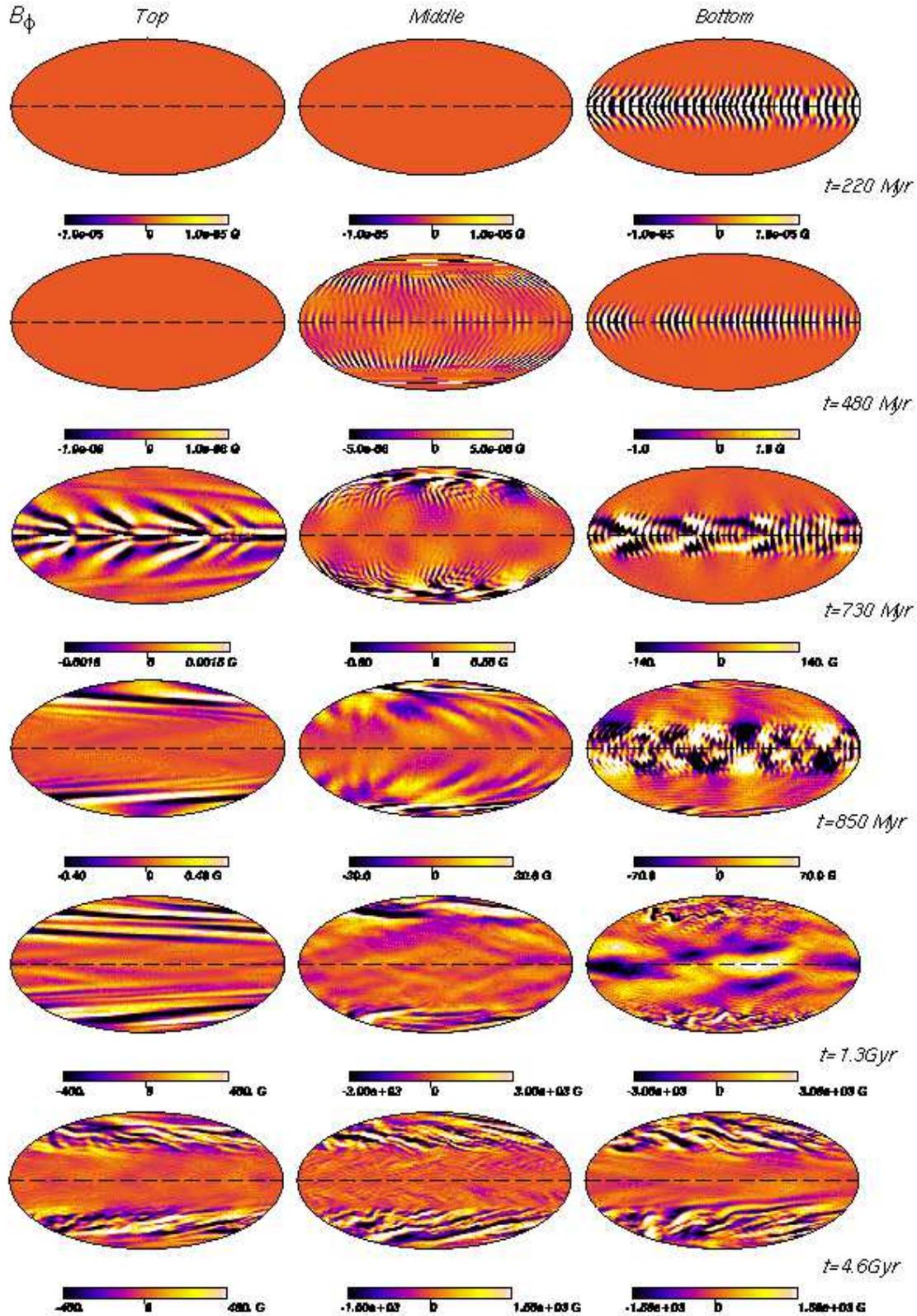}
\caption[]{\label{figShTac} Case III. Temporal evolution of the azimuthal magnetic field (from which we have substracted the $m=0$ component), shown in horizontal Mollweide projection  as a function of depth
(respectively from left to right: $r=0.7, \, 0.55$ and $0.37 R_{\odot}$), at various times in the evolution as indicated in solar equivalent units. First, the high-$m$ instability of the poloidal field appears near the equator at the bottom of the domain, and later the low-$m$ instability of the toroidal field occurs at all depths near the poles. Note the remarkably clean $m=1$ structure at the top in the frames for 0.85 and 1.3 Gyr. Between $\sim$ 500 Myr and 850 Myr there is a clear coexsistence of both instabilities in the simulation.}                  
\end{figure*}

\section{Instabilities} \label{instab}
Our three-dimensional simulations reveal another interesting 
and fundamental aspect of the problem, namely the occurrence of non-axisymmetric MHD instabilities in the radiation zone.  
These have been studied in detail through linear analysis by Tayler and collaborators (Markey \& Tayler 1973, 1974; Tayler 1980; Pitts \& Tayler 1985) and by Wright (1973); an excellent review of the subject has been written by Spruit (1999). Their main finding was that purely poloidal or purely toroidal fields are unstable to non-axisymmetric perturbations, and that stable configurations must therefore be of mixed poloidal/toroidal type, with comparable field strength.  Since then, such configurations have indeed been identified in recent numerical simulations performed by Braithwaite \& Spruit (2004, 2006) and Braithwaite \& Nordlund (2006).
Although our primary goal here was to examine the possibility that the solar tachocline could be confined by a fossil magnetic field, we feel that our calculations have also a broader scope, and that they may contribute to better understanding the behavior of stellar magnetic fields. We shall concentrate here on case III; cases I and II show similar behavior.
 
A useful quantitative diagnostic is provided by Fig.~\ref{figKEMETac}, which describes the temporal evolution of various components of the kinetic (KE) and magnetic (ME) energies, averaged over the whole computational domain. 
PKE \& PME designate respectively the mean (axisymmetric) poloidal 
components of KE and ME, TKE \& TME their mean toroidal components, and FKE \& FME their non-axisymmetric components 
(see Brun et al. 2004 for their analytic expressions).  
In Fig.~\ref{figShTac} we display the  maps, in Mollweide projection, of the non-axisymmetric part of the toroidal field, at  different depths and epochs; they will serve to locate and to characterize the instabilities.

\begin{figure}[!ht]
\center
\includegraphics[width=1.0\linewidth]{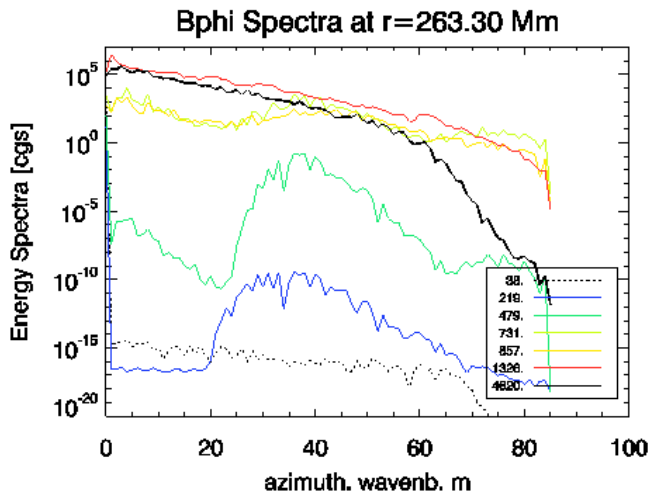}
\caption[]{\label{spectrum-bottom} Distribution over azimuthal wave-number $m$ (and summed of all $l >m$) of the magnetic energy in $B_\phi$, close to the bottom of the domain.
The two instability phases have distinct signatures: the first, due to the purely poloidal field configuration, occurs at $m > 20$ with a maximum around $m=40$, whereas the second, caused by the toroidal field,  starts around 450 Myr, and involves low $m$ wave-numbers. At about  1.3 Gyr, the highest wave-numbers are eroded through Ohmic diffusion.}                  
\end{figure}

>From the onset, starting from an extremely weak non-axisymmetric velocity field, FKE \& FME rise exponentially by many orders 
of magnitude at a fast pace (the e-folding time is $\approx 10$ Myr), and then they saturate around 700 Myr at a level 1000 times below the energy of the mean poloidal field (PME). As can be seen in 
Fig.~\ref{figShTac}, the fluctuations are located at low latitude and at the bottom of the domain; we conclude that it is the instability described by the pioneers quoted above, which occurs when the field is purely poloidal. Figure~\ref{spectrum-bottom}  displays the energy spectrum in $m$ of the fluctuations of the azimuthal field $B_\phi$ at the bottom of the domain: one sees clearly that the maximum growth occurs at rather high azimuthal wavenumber:  $m \approx 40$. This is 
in qualitative agreement with Wright (1973) and Markey \& Tayler (1974), who found that the growth-rate should steadily increase with $m$ in the non-dissipative case. Note in Fig.~\ref{figMC} that no mean toroidal field is produced in the unstable region, presumably because the magnetic helicity was initially zero. 

In the meanwhile, two bands of toroidal field of opposite polarity appeared in the upper part of the domain, one at mid latitude and the other close to the rotation axis, with opposite sign in each hemisphere. This field is  generated by the shearing of the poloidal field, which very soon encountered the differential rotation spreading down from  the convection zone.
At about 1.2 Gyr the energy TME, averaged over the whole domain, matches that of the mean poloidal field, meaning that locally the toroidal field can be much stronger than the poloidal field, since it  occupies a smaller volume. That toroidal field is unstable to low $m$ perturbations,  as can be seen both in Fig.~\ref{figShTac} and in the spectra of Fig.~\ref{spectrum-bottom}. 
Note that the mean toroidal field and the associated instability  draw their energy from the differential rotation, as indicated by the conspicuous inflection of the TKE curve in Fig.~\ref{figKEMETac}, at 1.1 Gyr.

Thereafter all energies but one slowly decline, 
at a rate which is controlled mainly by the Ohmic dissipation of the mean poloidal field (PME): in our rescaled solar units, the $e$-folding time would be $R^2/2  \pi^2 \eta \approx 2$ Gyr for a uniform motionless sphere (Cowling 1957), which is compatible with what we see here. The decay of that field is particularly  smooth, and 
there is no sign that it is regenerated or even affected by the instability. The mean toroidal field accompanies closely the decline of the mean poloidal field, 
and so do the non-axisymmetric components FKE and FME, with the kinetic energy in these perturbations being about twice their magnetic energy. 
The only exception to that overall decline is 
the mean toroidal kinetic energy TKE, which steadily increases as the differential rotation spreads deeper and to lower latitudes. 
The kinetic energy of the meridional flow (PKE) is the smallest of all energies; it is mostly concentrated toward the top of the domain and it remains almost constant during the whole evolution.


Recently Braithwaite \& Spruit (2004) studied Tayler's instability in a polytropic star, by means of 3-D numerical simulations. They found that a random initial magnetic field relaxes in a few Alfv\'en crossing times into a mixed poloidal/toroidal topology, by exciting a non-axisymmetric small scale instability. This stabilized field then slowly decays and migrates toward the surface. Our simulations differ here from Braithwaite \& Spruit in that we include rotation, initially impose  a large scale poloidal field, thus with zero helicity,  and that we apply differential rotation at the top the domain, which shears the slowly decaying poloidal field to produce a toroidal field, antisymmetrical with respect to the equator. Near the poles, our field is of the kind considered by Tayler (1973)
and it is known to give rise to non-axisymmetric instabilities at low azimuthal wave-number $m$; these are observed here in a fully developed regime, over a broad spectrum of scales, suggesting that they could mix the chemical elements in the radiation zone. However, none of the instabilities observed in our simulations seems able to regenerate
the mean ($m=0$) poloidal field from which we started the simulations. This indicates that our
simulations cannot be considered as dynamos, since the dynamo loop ($B_p \rightarrow B_{t} \rightarrow B_{p}$) is not fulfilled.
We stress however that this model does not apply to the Sun, since such a fossil field alone does not allow for a thin tachocline, as we have seen.  

\section{Discussion and conclusions}
We undertook these 3-D simulations primarily  to check whether a fossil magnetic field can prevent the radiative spread of the solar tachocline, 
as was advocated by Gough and McIntyre (1998). In all cases we have explored, the outcome is always that the poloidal field connects 
with the convection zone, and that it then imprints the differential rotation of that zone throughout the radiative interior, according to 
Ferraro's law, at latitudes higher than about 40$^\circ$ (see Figs. 4c-d). When the initial poloidal field threads into the convection zone, 
the process takes only a few Alfv\'en crossing times, as expected. When the field is confined in the radiation zone, the time required for this depends 
on how deeply it is buried. Nearly uniform rotation can only persist inside the outermost closed field line, hence at low latitude. Of course, 
a deeply buried field will not reach the convection zone by the age of the Sun, but then the tachocline will also have penetrated very far. 

The way the topology of the poloidal field enforces a given rotation profile was  elucidated by MacGregor and Charbonneau (1999), but they did not let the field diffuse into the convection zone. 
The stationary solutions built by Garaud (2000) resemble ours, especially in what she calls the low field case which applies to the field strength chosen here;  the main difference to our model is probably that her meridional circulation is driven by Ekman pumping. 

We believe that our simulations have captured the essence of the real problem, since they respect the similarity of  time-scales characterizing the 
two competing processes, namely the Ohmic diffusion time of the mean poloidal field and the Eddington-Sweet time which rules the ventilation and the spread of the 
tachocline. In Gough \& McIntyre's model, an important role is played by the magnetopause, namely that the boundary layer that separates the poloidal field in the uniformly rotating interior from the tachocline void of magnetic field: it is the seat of the production of the toroidal field. According to (\ref{smalldelta}) its thickness $\delta$ scales as $B_p^{-1/3}$, and to allow its existence in spite of the enhanced diffusivities, i.e. in order to ensure $\delta \ll r_{\rm top}$, we had to increase the initial poloidal field $B_p$ by 3 orders of 
magnitude with respect to what would be required in the Sun. This could still be insufficient, since in the cases discussed here the two boundary layers have similar thickness, 
whereas in Gough \& McIntyre's model the magnetopause is much thinner than the tachocline. To check this point we ran a simulation with enhanced thermal diffusivity and reduced Ohmic diffusivity, to increase the contrast $\Delta/\delta$ by a factor of 3: the outcome was the same, and the differences barely noticeable, except 
for slightly less regular poloidal field lines.

A more delicate point is whether the poloidal field lines actually penetrate into the convection zone, as assumed in most models, including ours, 
or whether they are deflected by the turbulent motions, which are probably structured in plumes. Realistic simulations of 
penetrative convection are required to see how a magnetic field, anchored in the deep interior, will behave there;  we intend to tackle this problem in the near future.

Since we used a three-dimensional code, we were able to observe the non-axisymmetric instabilities associated with our field configurations. In the early phase, they occur in a narrow equatorial belt at the base of the computational domain, and they are due to the 
purely poloidal field we have imposed initially. Later on, they are present at all depths near the poles, where the shear of the differential rotation keeps generating a strong toroidal field. Their properties agree with the predictions of Tayler and his collaborators, which were recently reviewed and completed by Spruit (1999, 2002). 

An important result of our simulations is that these instabilities do not interfere with the mean poloidal field; they are able to distort the `isogyres' (surfaces of constant angular velocity $\Omega$), thus affecting somewhat the production of the toroidal field, but  even there they seem to have limited impact. Further we do not see a dynamo process occuring
in our simulated radiative interior since the mean poloidal field is not regenerated and decays away.

Based on their model, and on the observed thinness of the tachocline, Gough and McIntyre (1998) concluded about the inevitability of a magnetic field in the Sun's radiative zone. We find here - but note the caveats above - that such a field seems unable to prevent the spread of that shear layer, and therefore that the thinness of the tachocline cannot prove or disprove the existence of a fossil field. This conclusion holds even when the aforementioned instabilities are taken into account, as in the new model presented by McIntyre (2006), since these instabilities do not affect the mean poloidal field.

Therefore other processes must be invoked to account for the thinness of the tachocline. Or one has to explain why there is no need for such a layer at the top of the radiation zone, 
and thus how uniform rotation is accomplished already at the base of the convection zone. A first step in that direction has been taken by 
Forg\'acs-Dajka and Petrovay (2000, 2002): they introduce the oscillating dynamo field in the region of convective penetration, and they succeed 
in smoothing out the differential rotation.  

But other possibilities are being actively explored. Talon and Charbonnel (2005; see also Charbonnel \& Talon 2005) have shown that such 
uniform rotation can be achieved with internal gravity waves emitted at the base of the convection zone, combined with meridional circulation and shear-induced turbulence in the bulk of the radiation zone. It is clear that in order to disentangle the theoretical models of the solar tachocline and the radiative interior currently 
available, a new class of helioseismic instruments has to be developed, in particular to study the deep solar interior. 
Such efforts have started in Europe (e.g. the DYNAMICS project, Turck-Chi\`eze et al. 2005).

\begin{acknowledgements}
We are grateful to N. Brummell and P. Charbon\-neau with whom we originally discussed aspects of this work.
We also acknowledge fruitful discussions with D. Gough, P.~Garaud, M.~McIntyre, H.~Spruit and J. Toomre. We thank the organizers
of the MSI program at the Newton Institute, N.~O. Weiss, R. Rossner and D. Hughes for their invitation during Fall 2004, 
where part of this work was completed. We wish to express our gratitude to the anonymous referee, whose constructive remarks led to substantial improvements of the manuscript. Finally we thank the French supercomputer centers CEA-CCRT and CNRS-IDRIS for their generous
time allocations, and CNRS (Programme National de Physique Stellaire) for its financial support.
\end{acknowledgements}

\end{document}